\documentclass[review]{elsarticle}

\usepackage{lineno,hyperref}
\modulolinenumbers[1]
\usepackage{epsfig}
\usepackage{amsmath}
\usepackage{booktabs}
\usepackage{hyperref}
\UseRawInputEncoding
\usepackage{gensymb}

\usepackage{subfigure}
\usepackage{geometry}
\usepackage{graphicx}
\usepackage{caption}

\usepackage{textcomp}
\usepackage{amssymb}
\usepackage{xcolor}
\usepackage{setspace}

\begin{document}

\begin{frontmatter}

\title{Measurement of undulator section wakefield at the SXFEL test facility}

\author[1,2]{He Liu }

\author[1,2]{Hanxiang Yang}

\author[1,2]{Nanshun Huang}

\author[1,4]{Liang Xu}

\author[3]{Zenggong Jiang}

\author[3]{Duan Gu}

\author[3]{Haixiao Deng\corref{cor1}}
\ead{denghaixiao@zjlab.org.cn}
\cortext[cor1]{Corresponding author}

\author[3]{Bo Liu}

\address[1]{Shanghai Institute of Applied Physics, Chinese Academy of Sciences, Shanghai 201800, China}

\address[2]{University of Chinese Academy of Sciences, Beijing 100049, China}

\address[3]{Shanghai Advanced Research Institute, Chinese Academy of Sciences, Shanghai 201210 , China}

\address[4]{School of Nuclear Science and Technology, Xi’an Jiaotong University, Xi’an 710049, China}

\begin{abstract}
In free electron laser facilities, almost every kind of device will generate wakefield when an electron beam passes through it. Most of the wakefields are undesired and have a negative effect on the electron beam, which means a decrease of FEL performance. As for the SXFEL test facility, the sophisticated layout and the cumulative effect of such a long undulator section lead to an obvious wakefield, which is strong enough that can not be ignored. Based on two deflecting cavities at the entrance and the exit of the undulator section with corresponding profile monitors, we measured the wakefield of the undulator section. In this paper, we give the theoretical and simulation results of resistive wall wakefields which agrees well with each other. In addition, the experimental and the simulation results of the overall undulator wakefield are given showing small difference. In order to explore the impact of this wakefield on FEL lasing, we give the simulation results of FEL with and without wakefield for comparison. There is almost no impact on 44 nm FEL in stage-1 of cascaded EEHG-HGHG mode, while the impact on 8.8 nm FEL in stage-2 becomes critical and it decreases the pulse energy and peak power by 42\% and 27\% and broadens the bandwidth. 
\end{abstract}

\begin{keyword}
wakefield measurement \sep wakefield simulation \sep resistive wall wakefield \sep geometry wakefield \sep wakefield impact on FEL 
\end{keyword}

\end{frontmatter}

\renewcommand{\baselinestretch}{2.0}
\section{Introduction}\label{1}

Free electron laser (FEL) is one of the most advanced light sources around the world due to its advantages of high power, high coherence and short wavelength, supporting many cutting-edge researches in various fields, including chemical dynamics, molecular biology science, material analysis and so on\cite{barletta2010free}. In order to meet the demands of more advanced research, fully coherent X-ray FEL has been developed\cite{huang2021features,feng2018review}. Shanghai soft X-ray free electron laser (SXFEL) test facility is the first coherent X-ray light source in China and one of the eight completed X-ray FEL facilities around the world. It mainly consists of a linear accelerator (linac) which accelerates 600 pC electron beam to 840 MeV, and an FEL amplifier which can support many kinds of schemes, like HGHG, EEHG and cascaded 
schemes\cite{yan2021self}. SXFEL has achieved lasing at 8.8 nm.

Consider a bunch of electrons moving in a pipe with resistive wall or rough inner surface, or passing through a discontinuous structure, the longitudinal and transverse wakefields\cite{craievich2010short} could be generated and change the behaviors of the electron beam. Corrugated pipes are designed to generate special wakefields\cite{novokhatski2015wakefield,h.x.deng2014review} to extend the bandwidth\cite{zagorodnov2016corrugated}, linearize the longitudinal beam phase space\cite{deng2014experimental}, measure the beam temporal profile\cite{bettoni2016temporal}, support the fresh-slice multicolor scheme\cite{lutman2016fresh} and mostly, remove the energy chirp\cite{emma2014experimental,bane2012corrugated,bane2016dechirper,zhang2014design}. In addition, the wakefields could also be generated in other parts of an FEL facility, which bring about unexpected influence. Under usual circumstance, the energy loss induced by the longitudinal wakefield of a single device (not designed for wakefield utilize) is small enough compared with the energy of electron that not taken into account. However, the length of the undulator section in the SXFEL test facility is over 100m and there is a number of devices with sophisticated structure. As a result of cumulative effect, the longitudinal passive wakefield of the whole undulator section could be strong enough and should not be neglected. As well researched, the longitudinal wakefield could lead to the head-to-tail energy deviation of the electron beam\cite{bane2015using}. If the electron beam is off the target energy, the resonance relation\cite{huang2007a} could not be satisfied which means a degradation of FEL or even failing in lasing. The electron beam can also generate a strong wakefield in the linac\cite{bane2003measurement} but its energy feedback system is able to monitor and compensate the energy deviation\cite{song2016wakefield}. Therefore the measurement of undulator section wakefield is of great significance for FEL lasing especially for the cascaded schemes of the SXFEL test facility. 

The wakefields of the undulator section mainly consists of two parts: one is the resistive wall wakefield\cite{stupakov2015resistive,podobedov2009resistive} of vacuum chambers and other resistive pipers, another is the geometry wakefield of discontinuous structures. The inner surface of all the devices are considered to be smooth enough that the surface roughness wakefield is not counted. The theoretical formulas of the resistive wall wakefields have been well researched and the simulation results of vacuum chambers are given compared with the theoretical results showing great agreements. We give the simulation wakefield result of overall undulator section, which also agrees well with the experimental result. 

In this article, we initially introduce the SXFEL test facility. Then we analyze the source of the wakefield in the undulator section of the SXFEL test facility. Subsequently, we give simulations of the wakefield using the wakefield solver, ECHO2D\cite{zagorodnov2015calculation}. Following that is the experimental result of the wakefield measurement which agrees well with the simulation result. Ultimately, the analysis towards the beam energy loss impact on FEL performance based on simulation is given.

\section{SXFEL test facility}

SXFEL test facility is designed for external seed laser based on harmonic up-conversion scheme\cite{yan2021self}. The schematic layout of the SXFEL test facility is shown in Fig.1. The electrons are generated by a photo-cathode RF gun and could be accelerated to 840 MeV by the linac section which contains S-band and C-band accelerators\cite{yan2021self}. After accelerated the electron beam enters the undulator section,  which consists of two stages of modulators, dispersive chicanes and undulators. Either the first stage or the second stage can serve as a high gain harmonic generation (HGHG) scheme independently, and obviously, cascaded HGHG is also feasible. Before the first stage there is an extra group of modulator (M0) and dispersive chicane (DS0), which makes the echo-enabled harmonic generation (EEHG) available working with the first stage. In this case, SXFEL test facility can lase in single-stage HGHG, single-stage EEHG, cascaded HGHG and cascaded EEHG-HGHG schemes. SXFEL test facility has achieved providing 8.8 nm FEL operating in cascaded EEHG-HGHG scheme with a 264 nm pulsed laser as the initial signal. 

As is illustrated in Fig.1, there are two deflecting cavities at the entrance and the exit of the undulator section\cite{song2017deflecting} with corresponding profile monitors. The first deflector is a C-band deflector and the second deflector is an X-band deflector. The time resolution of these two deflectors could reach around 30 fs and 6 fs, which are designed to diagnose the time-resolved electron beam energy and longitudinal profile of FEL radiation. Based on these devices, we can measure the longitudinal phase space of an electron beam at the entrance or the exit of the undulator section, sequentially the energy loss is able to be calculated. This offers a condition to our experiment of the overall undulator section wakefield measurement.

\begin{figure}
\centering
\includegraphics[width=\textwidth]{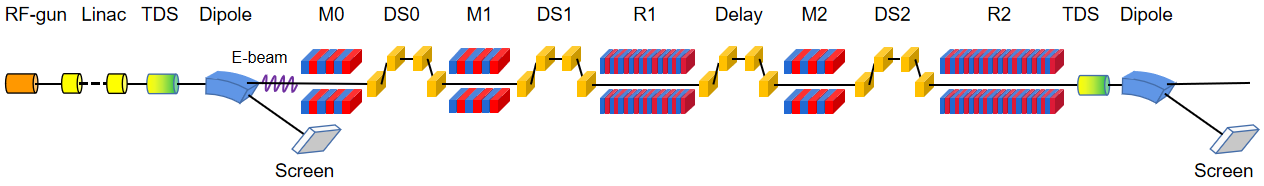}
\caption{Layout of the SXFEL test facility.}
\end{figure}

\section{Wakefield calculations}

The vacuum chambers and other pipes in the undulator section of the SXFEL test facility are non-superconducting but mainly made of aluminum with the room temperature conductivity $\sigma =3.7\times10^7  (\Omega\cdot m)^{-1}$\cite{stupakov2015resistive}, which means the resistive wall wakefields are generated during the whole undulator section. The undulators are out vacuum planar undulators thus the wakefield in an undulator is counted as the resistive wall wakefield in the resistance chamber. In addition, the structural variation also leads to another kind of wakefield named geometry wakefield. The contribution of geometry wakefield mainly comes from flanges, corrugated pipes, step-in and step-out structures. The internal surface of all the devices are considered to be smooth enough that the surface roughness wakefield is not counted\cite{song2017wakefields}. In order to calculate the wakefield of overall undulator section, we used the wakefield solver, ECHO2D, which supports both resistive wall wakefield and geometry wakefield calculations. For comparison, we calculated the resistive wall wakefields of vacuums chamber with formula\cite{stupakov2015resistive}:

\[Z(k) = \frac{Z_0}{2\pi a} \left( \frac{1}{\xi(k)}-\frac{ika}{2} \right) \]
\[w(s) = \frac{c}{2\pi} \int_{-\infty}^{\infty} Z(k)e^{-iks}dk\]

where $Z_0$ is the free space impedance, $Z_0=120\pi \Omega$, a is the diagram of the chamber, k is frequency and c is the speed of light. We calculated the resistive wall wakefields of round chambers with diameters of 6 mm and 16 mm and the beam current we used for both simulation and theoretical calculation was measured in the experiment. The current profile we used is shown in Fig.2 together with the wake potential calculated by ECHO2D and formulas (The bunch head is on the left). The bunch charge is 500 pC and the bunch length is 1.3 ps (FWHM). As is illustrated in Fig.2 (a) and (b), the theoretical and simulation results agree well with each other. It is obvious that the wake potential generated by a $\Phi$6 vacuum chamber is more than twice the wake potential generated by a $\Phi$16 vacuum chamber, which is reasonable according to the formulas mentioned above.

\begin{figure}[htbp]
\centering
\subfigure[]{
\begin{minipage}{0.3\textwidth}
\centering
\includegraphics[width=4cm]{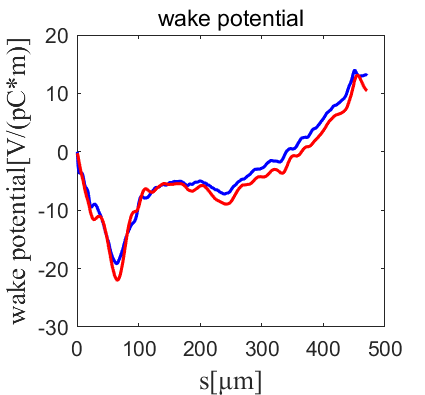}
\end{minipage}
}
\subfigure[]{
\begin{minipage}{0.3\textwidth}
\centering
\includegraphics[width=4cm]{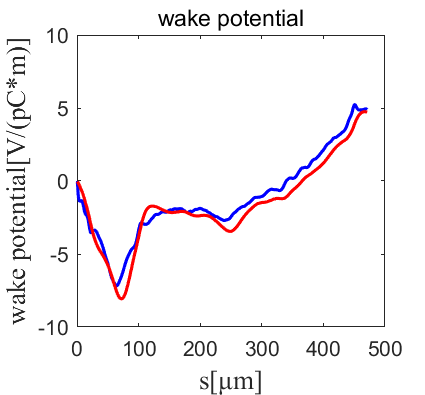}
\end{minipage}
}
\subfigure[]{
\begin{minipage}{0.3\textwidth}
\centering
\includegraphics[width=4cm]{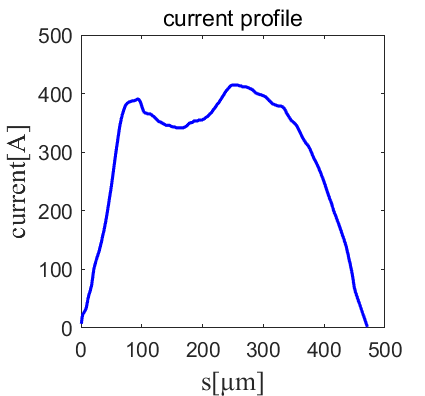}
\end{minipage}
}
\caption{\label{fig}(a) and (b) are the theoretical and simulation wake potentials of Φ6 and Φ16 vacuum chambers, the blue lines are the results of ECHO2D and the red lines are the theoretical results.(c) is the current profile.}
\end{figure}

For simplifying the calculation and saving time, we divide the undulator section into three modules: vacuum chambers with diameters of 16 mm and 6 mm, and inter sections between two undulators. The total length of $\Phi$16 and $\Phi$6 vacuum chambers are 35 m and 74.439 m respectively, and the total number of inner sections is 7. Thus we just need to calculate the wakefield of 10mm vacuum chambers and one inner section, the overall wakefield is the sum of every kind of the three modules multiplies by their corresponding multiples. We calculated the wake potential of these three models and the energy loss of the overall undulator section. The total energy loss of the real bunch and the Gaussian bunch ($\sigma$=120 \textmu m) calculated by ECHO2D are shown in Fig.3.

\begin{figure}[htbp]
\centering
\subfigure[]{
\begin{minipage}{0.48\textwidth}
\centering
\includegraphics[width=6cm]{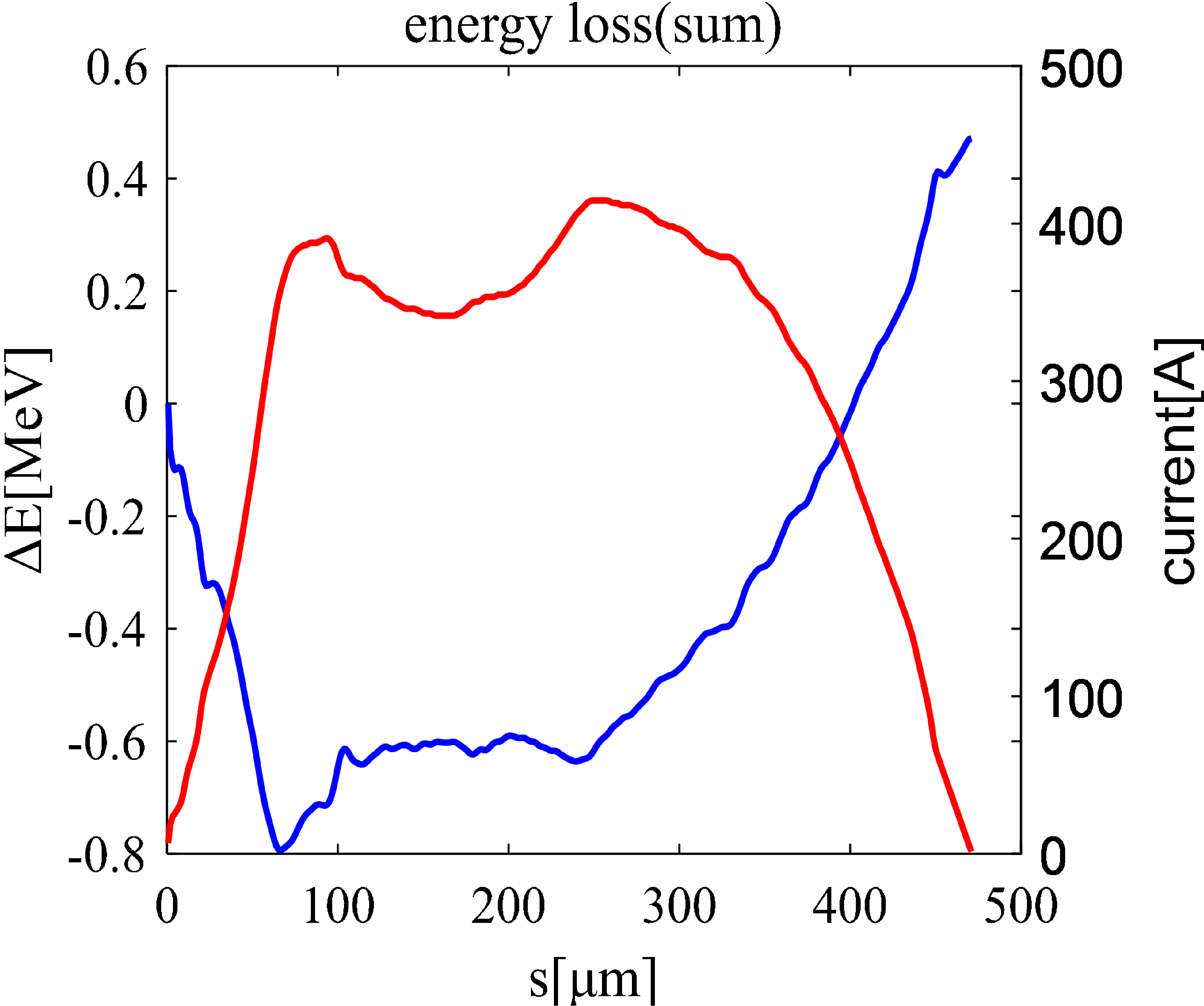}
\end{minipage}
}
\subfigure[]{
\begin{minipage}{0.48\textwidth}
\centering
\includegraphics[width=6cm]{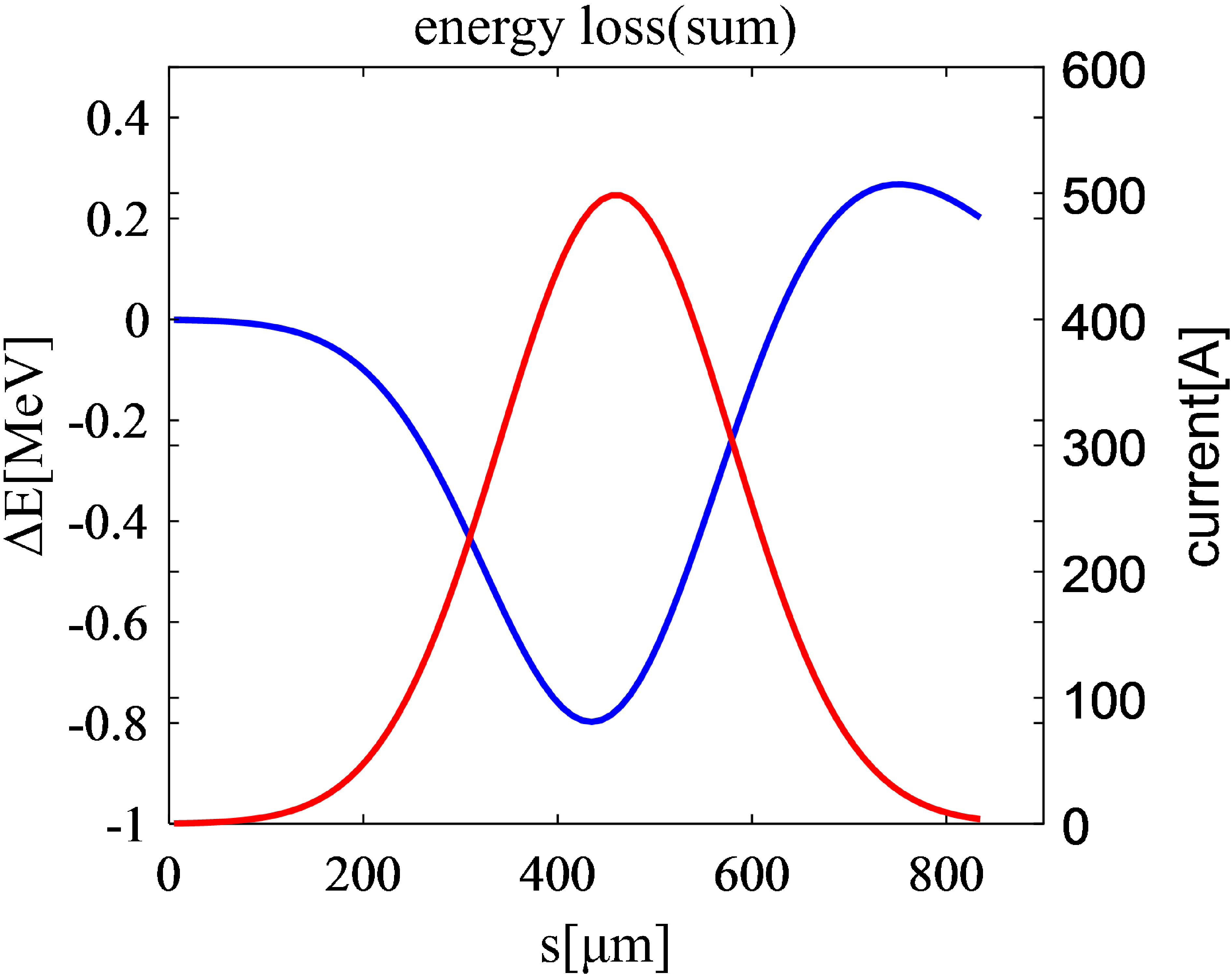}
\end{minipage}
}
\caption{\label{fig}The energy losses of the overall undulator section calculated based on the real bunch(a) and the Gaussian bunch(b). The red lines show the corresponding current profile.}
\end{figure}

The inner sections containing all the discontinuous structures generate geometry wakefields and resistive wall wakefields simultaneously, which contributed the most energy loss in the undulator section. As shown in Fig.3, the energy loss calculated based on a Gaussian bunch agrees well with that calculated based on the real bunch, which makes the simulations reliable. The mean beam energy loss(the real bunch) of the overall undulator section is 0.48 MeV, and the maximal beam energy loss is 0.79 MeV. 

\section{Wakefield measurement}

We measured the wakefield of the undulator section through deflecting cavities and corresponding profile monitors. What should be emphasized is that the undulators were turned off while the experiment was carrying on. Thus the impact of the undulators could be ignored. We obtained the longitudinal phase space of the electron beam at either the entrance or the exit of the undulator section. Fig.4 shows two of these corresponding measuring results . 

\begin{figure}[htbp]
\centering
\subfigure[]{
\begin{minipage}{0.48\textwidth}
\centering
\includegraphics[width=6cm]{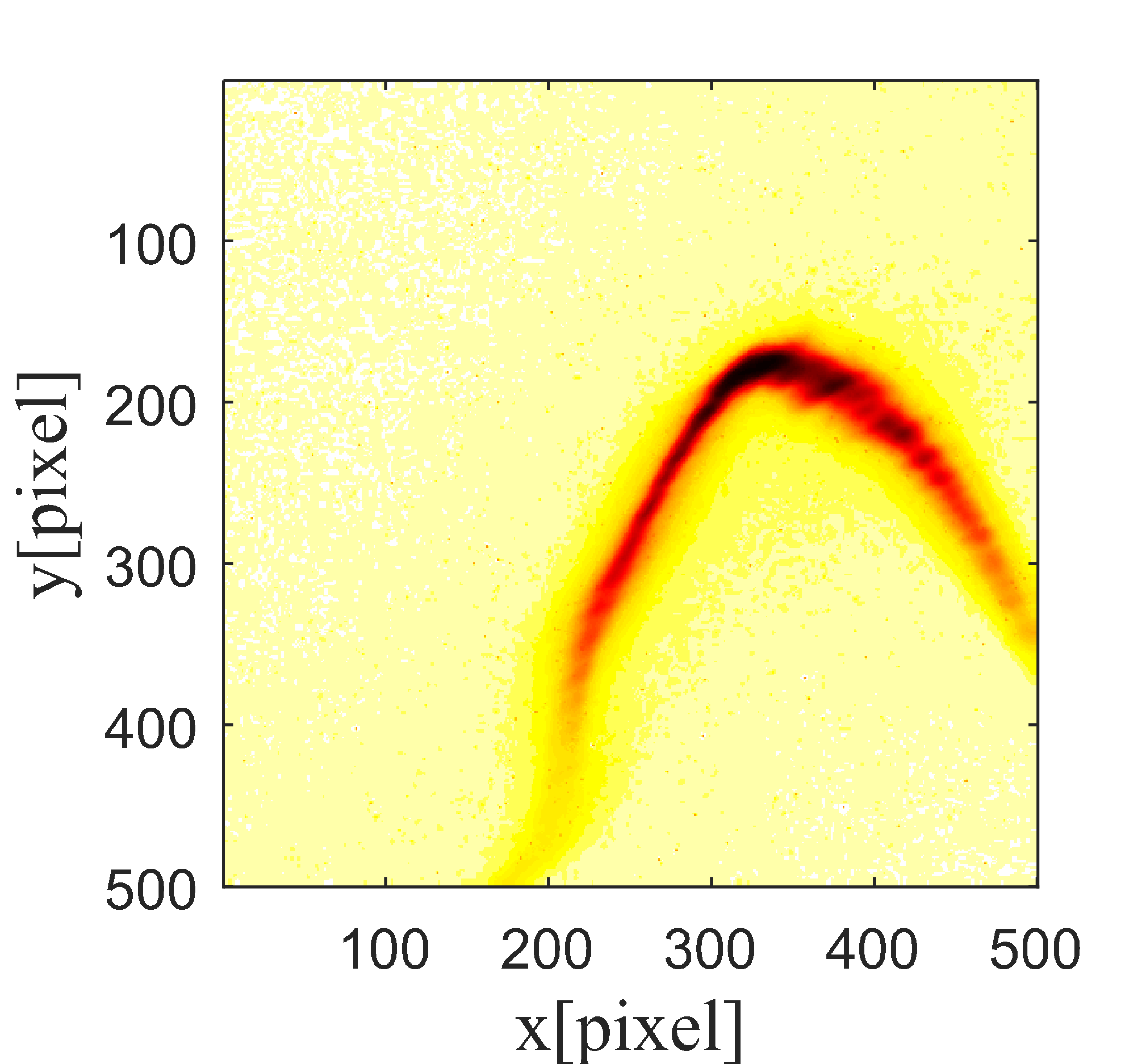}
\end{minipage}
}
\subfigure[]{
\begin{minipage}{0.48\textwidth}
\centering
\includegraphics[width=6cm]{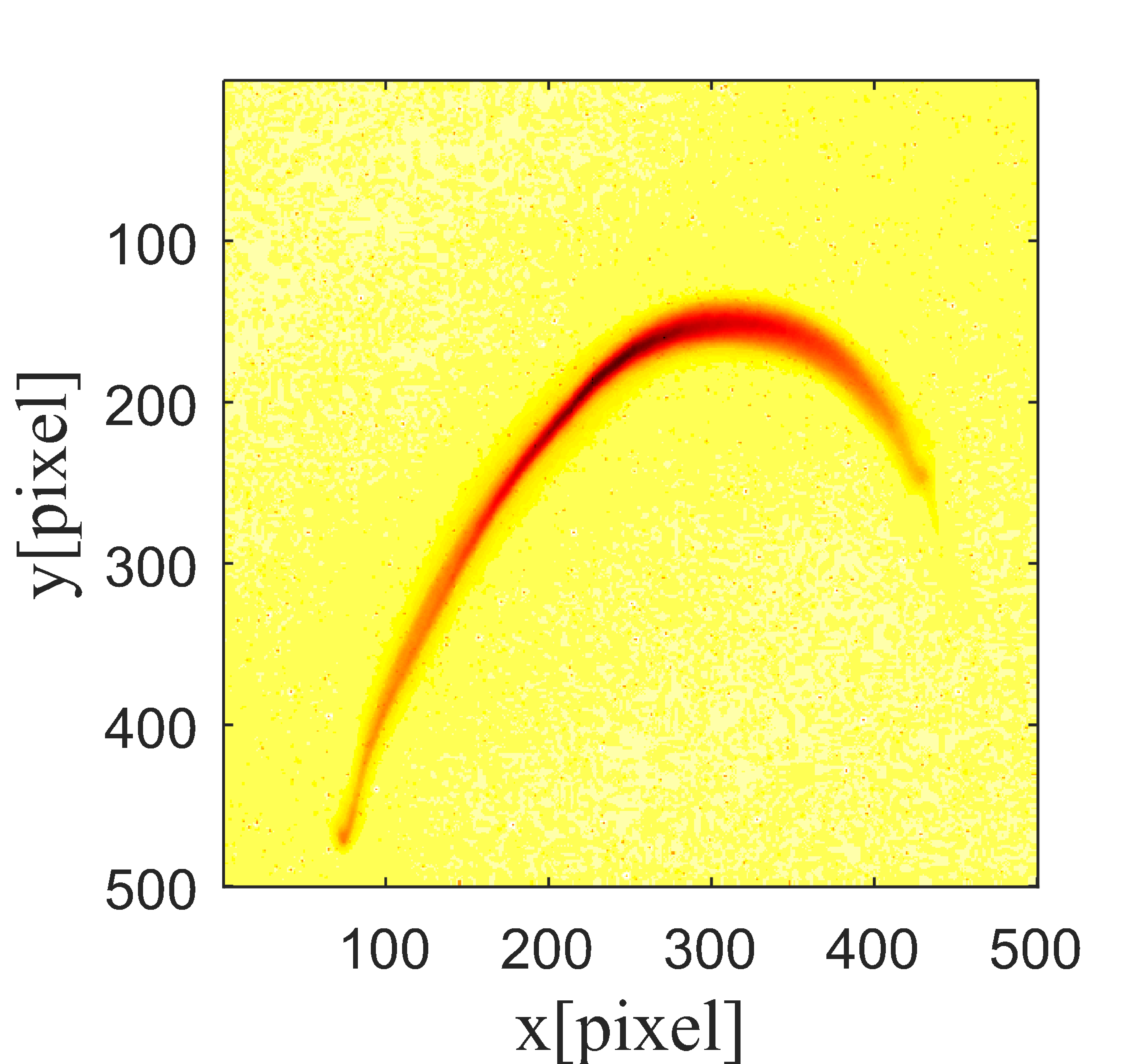}
\end{minipage}
}
\subfigure[]{
\begin{minipage}{0.48\textwidth}
\centering
\includegraphics[width=6cm]{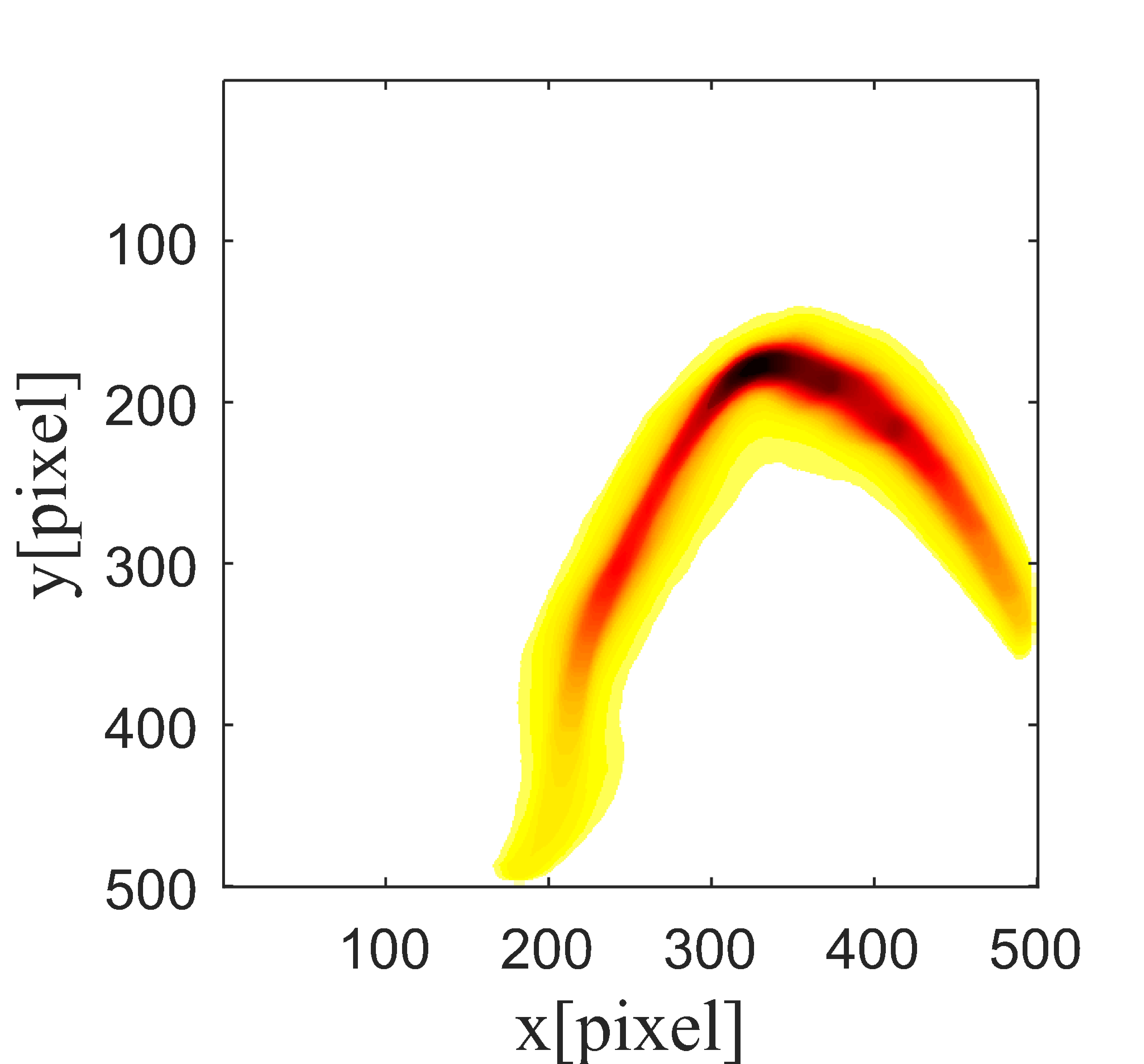}
\end{minipage}
}
\subfigure[]{
\begin{minipage}{0.48\textwidth}
\centering
\includegraphics[width=6cm]{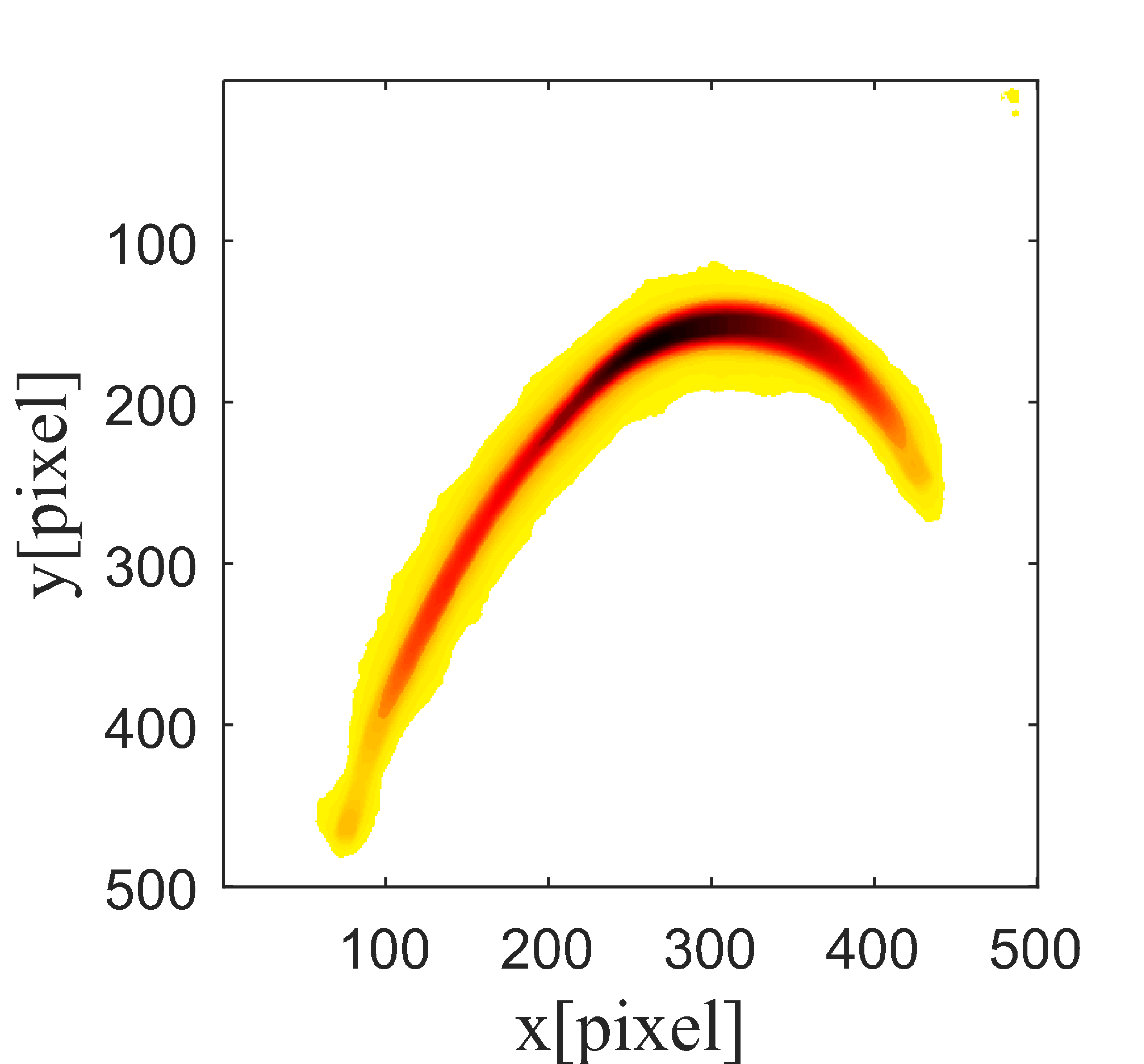}
\end{minipage}
}
\caption{\label{fig}The longitudinal phase space of the electron beam measured at the entrance (a) and the exit (b) of the undulator section. (c) and (d) are the results after the noise reduction and the smoothing processing of (a) and (b).  }
\end{figure}

As is shown in Fig.4, there are a lot of bottom noise and burr in the measuring results, thus we carried on some corresponding processes to remove these noise and burr. The data matrices of the measuring results are 500$\times$500 matrices. Each element in the matrices represents the number of electrons whose column number (x-axis) and row number (y-axis) respectively represent the longitudinal position and the energy of these electrons. The units of the longitudinal position and the energy are both pixels and need to be calibrated. In order to calibrate the x-axis, we cut off the blank columns and calculated the current profile of each bunch by summing the data matrices along the X direction. Then we stretched or compressed the current profile matrices by interpolation method to an average matrix length value and calibrated their x-axis according to the FWHM bunch length measured in the experiment. In the same way, the x-axis of the original data matrices could be calibrated. Actually calibrating the y-axis is not necessary in this experiment, because what we care about is not the absolute value of the electron energy but the energy loss. We could always calculate the weighted average pixel difference of y-axis of two bunches and convert it to the energy loss in accordance with a pixel-to-energy conversion factor (25 keV/pixel). Due to some unknown noises and engineering errors, the images were jittering on the screen that we could not collect aligned images. This might lead to a nonzero value at the beginning of the energy loss, so we added a corresponding vertical translation to guarantee all the energy loss matrices begin with zero.

One thing that should be noted is that this measurement is an interception measurement, which means the longitudinal phase space we got at both ends of the undulator section were not from the same electron beam. To solve this problem we tried to find the most similar bunches. As long as the valid data we collect is enough that we could find several groups of current profile having satisfying similarities, the longitudinal phase space measured at the entrance and the exit of the undulator section can be seemed as measured from the same electron beam. We respectively measured 200 groups of data at the entrance and the exit. In order to find the most similar electron beams, we wrote a code to calculate the absolute difference between every current profile of two bunches separately measured at the both ends. We found the most similar 16 groups whose similarities of current profiles are all higher than 96.5\% and the current profile of the most similar group is shown in Fig.5(a). To show the importance of finding the similar electron beams, we put the current profile of a group of bunches in Fig.5(b) which shows a bad similarity. We compared the energy losses of these chosen groups and picked up ten of them which show great performances. These ten groups were prepared for the energy loss calculation. Fig.5 (c) and (d) show the energy losses calculated based on three of the ten groups and the average energy loss of all the picked groups. 

\begin{figure}[htbp]
\centering
\subfigure[]{
\begin{minipage}{0.48\textwidth}
\centering
\includegraphics[width=6cm]{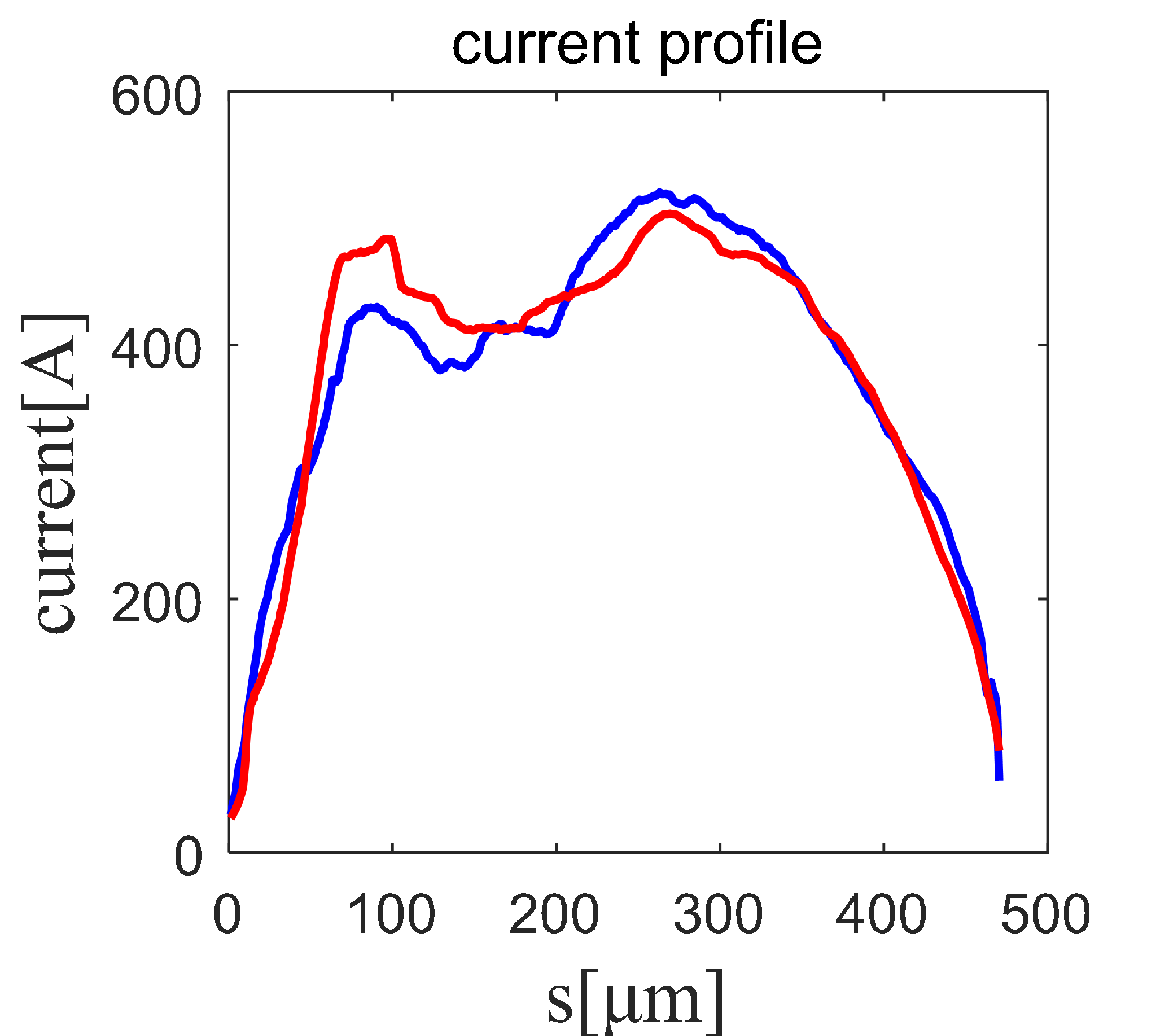}
\end{minipage}
}
\subfigure[]{
\begin{minipage}{0.48\textwidth}
\centering
\includegraphics[width=6cm]{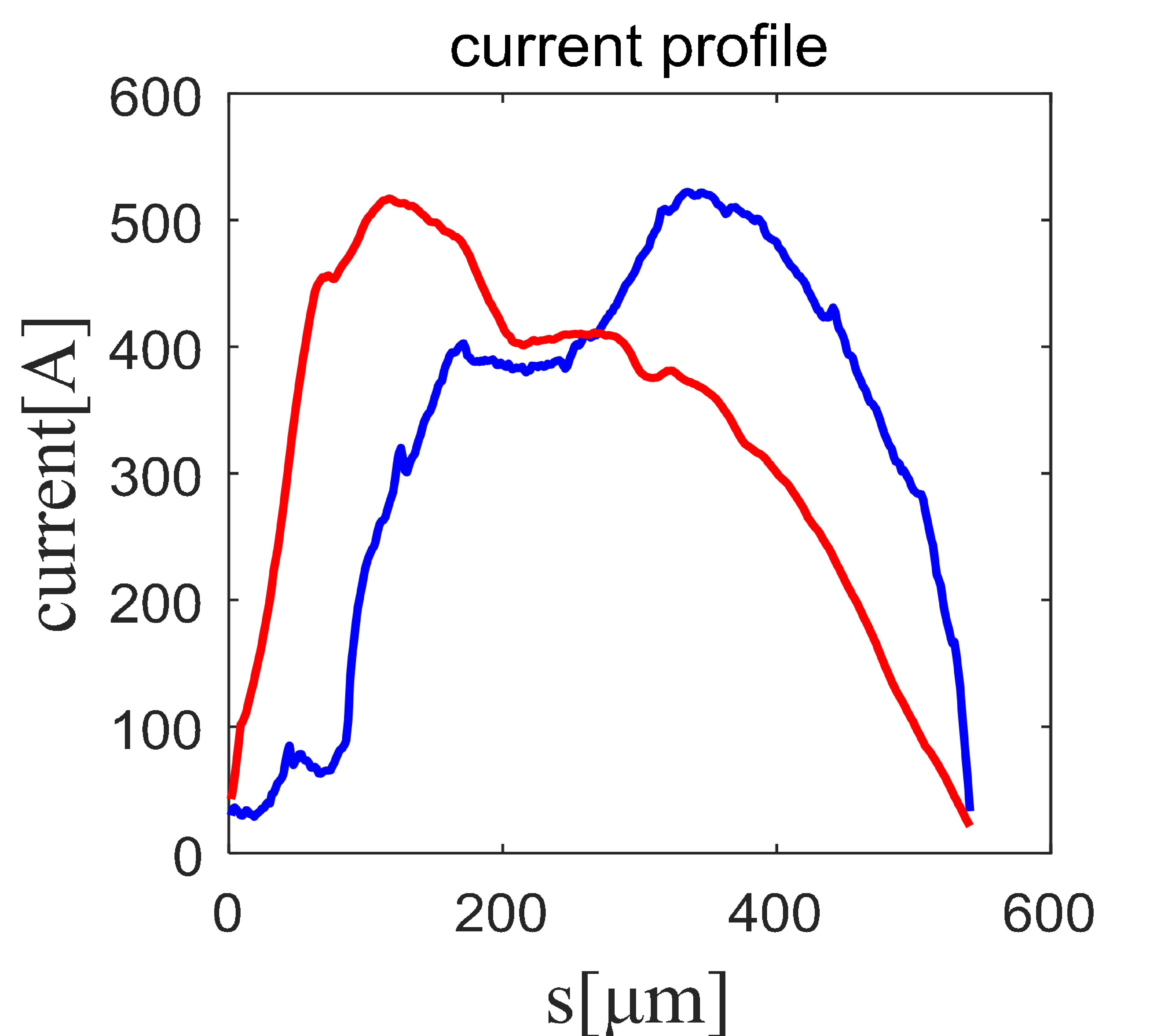}
\end{minipage}
}
\subfigure[]{
\begin{minipage}{0.48\textwidth}
\centering
\includegraphics[width=6cm]{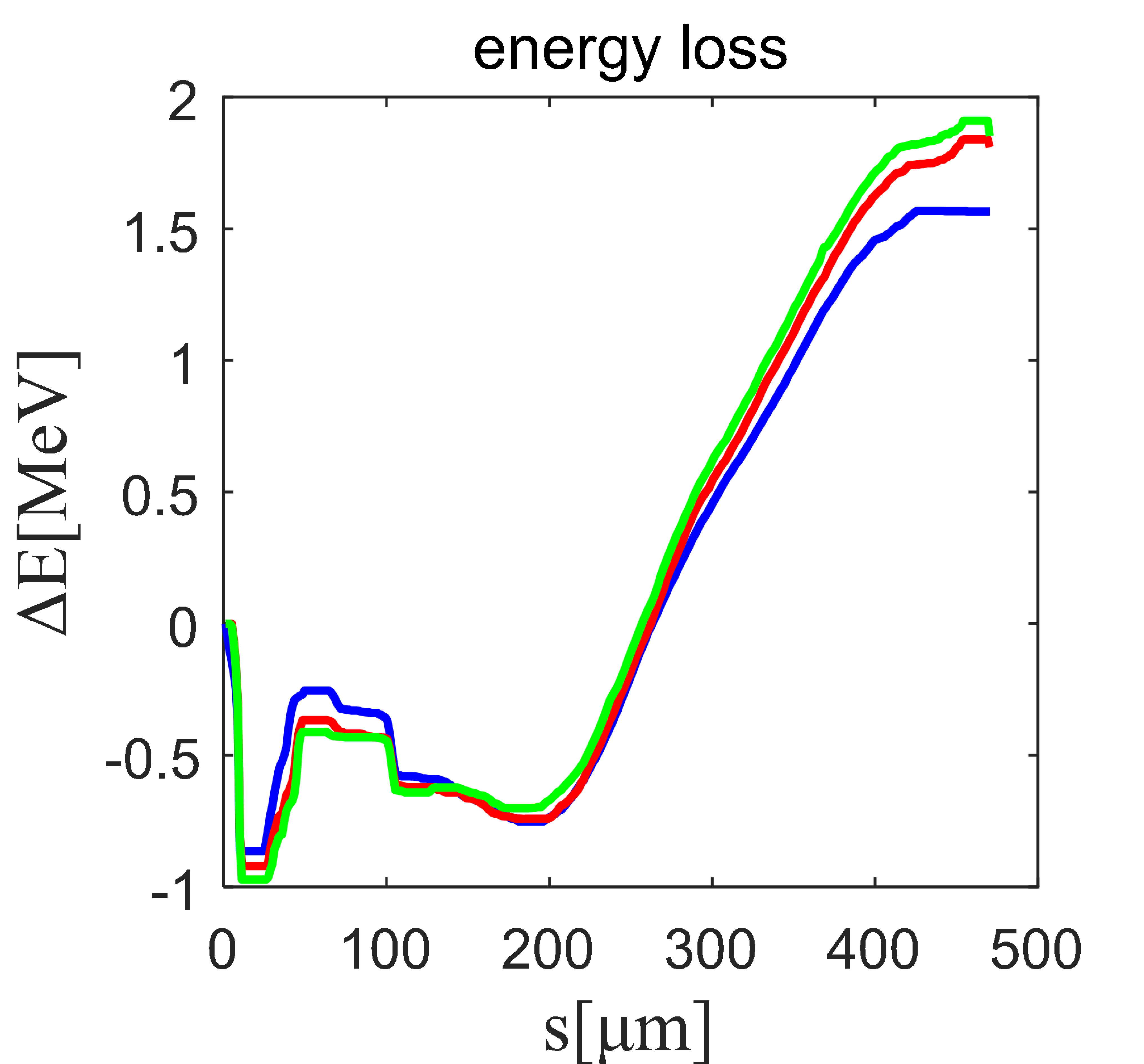}
\end{minipage}
}
\subfigure[]{
\begin{minipage}{0.48\textwidth}
\centering
\includegraphics[width=6cm]{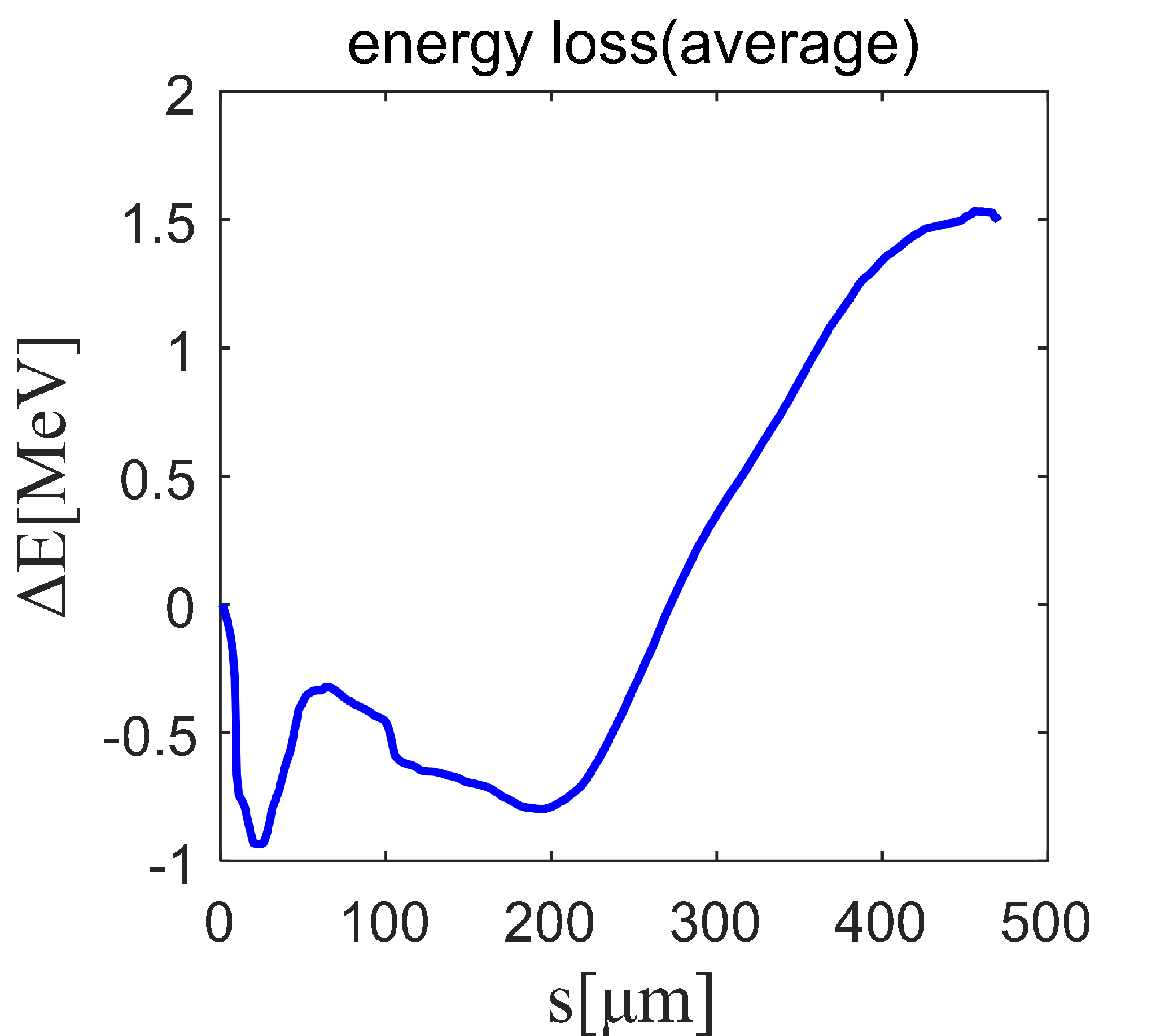}
\end{minipage}
}
\caption{\label{fig}(a) and (b) are the current profiles of two groups of bunches with high and low similarity. (c) is the energy loss of the three groups picked. (d) is the average energy loss of all of the groups available.}
\end{figure}

In Fig.5 (c) we only displayed three representative results but the other results also show the similar curves actually. Each curve of energy loss has two horns. This is reasonable due to the characteristics of current profile. The maximal beam energy loss is 0.93 MeV which is close to the simulation result but a little bit higher. This is predictable due to the sophisticated layout of the undulator section. In our calculations the geometry input of simulation was simplified, but there are some subtle geometry structures of some devices has been ignored which could generate geometry wakefields. There is a sharp rise in the latter part of each experimental energy loss result, which is different from the theoretical results (not that sharp). This difference might came from the transverse wakefields since the electron beam possibly move off the axis. Regardless of the small difference, the similar curves and the closed energy loss made the simulation and experimental results reliable.

A meaningful work is to calculate the wake function of the undulator section. Since the wake function is only related to the structure, material and layout of devices, it is fixed no matter how the current profile changed. The wake potential of the undulator section could always be calculated easily according to the current profile through convolution once the wake function is clear. We tried three methods to 
retrieve the wake function according to the wake potential measured in this experiment: deconvolution method, Fourier transform method and matrix method. The first method is only suitable for theoretical wake potential of a Gaussian bunch due to its iteration calculation, while the second method even cannot give a reasonable result. In order to avoid the iteration calculation resulting in mistakes, we changed the convolution to matrix multiplication: $CW_F=W_P$. $W_F$ and $W_P$ are wake function and wake potential. C is a square matrix consists of 1D current profile matrices offset with each other. In this matrix we add zeros for filling the blank. Thus we could obtain wake function through $W_F=C^{-1}W_P$. This method could retrieve the resistive wall wake function calculated by theoretical wake potential of both the Gaussian bunch and the real bunch, which showed perfect fit with original theoretical wake function. Further more, the results showed perfect performance even if we added noise to the current profile. Unfortunately the results became terrible if we smoothed the current profile and the original wake function after adding noise, which actually happened in data process of the experiment. It is predictable that the third method failed to give the correct wake function of the undulator section based on experimental results. Nevertheless, it is still a meaningful exploration. We hope that more accurate measurements could be achieved in our next experiment at the SXFEL user facility. By measuring with the same deflecting cavity, the noise could be significantly reduced and the matrix method might be successfully used to retrieve the wake function.

\section{Wakefield impact on FEL lasing}

The wakefield of the undulator section has been simulated and measured and one more important thing which is of great interest is how the wakefield impacts on FEL lasing. We adopted 3D time-dependent numerical simulation codes GENESIS\cite{reiche1999genesis} to simulate the FEL lasing with and without wakefield of the undulator section to show this impact through comparison. 

A problem that should be solved before simulation is that the experimental result contains the wakefields of the devices after the last undulator, which has no impact on FEL lasing but can not be removed. In addition, some useless drift sections are usually ignored in GENESIS simulation while the wakefields were also counted in the experimental result, which means it is not accurate if we use the experimental result for GENESIS simulation straightly. As for simulation results of ECHO2D, we can only count the requisite wakefields by piecewise calculation. Under this circumstance, we compared the ECHO2D simulation results of two different cases. One is the wakefield of the overall undulator section as mentioned above and another only contains the specific sections needed in GENESIS simulation. We calculated the ratio of these two cases and considered it was the same for experimental results. Thus we could obtain a ``ratio result'' of the experiment by multiply the ratio, which is equivalent to the experimental wakefields of the requisite sections. In this way we could compare the wakefield impact on FEL lasing simulated with three different wakefield inputs: segment simulated wakefield (the most accurate one), average ``ratio simulated wakefield'' and average ``ratio experimental wakefield''. As long as the results of the first two cases mentioned above agree well with each other, the average ``ratio experimental wakefield'' could be considered as a representation of the realistic experimental wakefield which makes it reasonable to be a wakefield input of GENESIS simulation. This means the impact on FEL lasing simulated based on the average ``ratio experimental wakefield'' is reliable and the problem presented at the beginning of this paragraph has been solved. Main parameters used in GENESIS simulation are listed in Table.1. The periods of three modulators and two radiators are 8 cm, 8 cm, 4 cm and 4 cm, 2.35 cm.

\begin{table}[!hbt]
	\centering
	\caption{Main simulation parameters of SXFEL test facility operating at cascaded EEHG-HGHG  mode}
	\begin{tabular}{p{6cm}p{3.5cm}}
		\toprule
		\textbf{Specifications}	        & \textbf{Electron beam}          \\
		\midrule
		Energy            			    & {840} {MeV}		      \\ 
		Slice energy spread             & {30} {keV}               \\ 
		Normalized emittance            & {1.5} {mm$\cdot$mrad}   	      \\ 
		Peak current                    & {415} {A}                \\
		Bunch length                    & {1.3} {ps} (FWHM)         \\
		\toprule
		\textbf{Specifications}       	& \textbf{Seed laser}             \\
		\midrule
		Peak power of seed1             & {45.9} {MW} (Gaussian)   \\ 
		Peak power of seed2             & {183.6} {MW} (Gaussian)  \\ 
		Duration			            & {160} {fs} (FWHM)        \\ 
		Wavelength                      & {264} {nm}      	      \\
		Rayleigh length                 & {47.6} {m}               \\
		\bottomrule
	\end{tabular}
	\label{tab:sim}
\end{table}

The simulation results of FEL lasing without and with three different kinds of wakefields mentioned above are listed in Table.2 and the FEL performance under the impact of the average ``ratio experimental wakefield'' are illustrated in Fig.6. Comparing the data of ``ratio simulation'' and ``segment simulation'', we could find the differences between them are quite small, which means the results of ``ratio measurement'' are reliable. The pulse energy and peak power of FEL without and with ``ratio measurement'' wakefield at the end of stage-1 are 29.0 \textmu J, 26.7 \textmu J and 286.8 MW, 272.0 MW. This means the wakefield impact on FEL is slight in stage-1, but the impact become remarkable in stage-2 as listed in Table.2. The pulse energy decreases from 21.0 \textmu J to 12.1 \textmu J and the peak power decreases from 219.7 MW to 160.2 MW.  The attenuation of pulse energy caused by wakefield is even nearly half of the case without wakefield and the peak power decreases by 27\%. A 23\% bandwidth broadening is also generated in stage-2 due to the wakefield. Actually pulse energy and peak power measured straightly at the end of stage-2 on the SXFEL test facility are more than 10 \textmu J and 100 MW, which makes the simulation results reliable. 

\begin{table}[!hbt]
	\centering
	\caption{The simulation results of FEL lasing with different kinds of wakefields}
	\begin{tabular}{lcccc}
    \toprule
	  & \multicolumn{1}{c}{\textbf{Pulse energy}} & \textbf{Peak power} & \textbf{Pulse length } & \textbf{Bandwidth } \\
	\midrule 
    \textbf{Without Wakefield} &                &          &          &         \\
            Stage1 EEHG        & 29.0 \textmu J & 286.8 MW & 100.5 fs & 0.103\% \\
            Stage2 HGHG        & 21.0 \textmu J & 219.7 MW & 105.7 fs & 0.019\% \\
    \textbf{Segment simulation}&                &          &          &         \\
            Stage1 EEHG        & 27.8 \textmu J & 286.6 MW & 95.2 fs  & 0.103\% \\
            Stage2 HGHG        & 16.9 \textmu J & 169.6 MW & 105.7 fs & 0.019\% \\
    \textbf{Ratio simulation}  &                &          &          &         \\
            Stage1 EEHG        & 26.3 \textmu J & 259.3 MW & 100.5 fs & 0.113\% \\
            Stage2 HGHG        & 14.3 \textmu J & 161.6 MW & 105.7 fs & 0.023\% \\
    \textbf{Ratio measurement} &                &          &          &         \\
            Stage1 EEHG        & 26.7 \textmu J & 272.0 MW & 95.2 fs  & 0.113\% \\
            Stage2 HGHG        & 12.1 \textmu J & 160.2 MW & 84.6 fs  & 0.024\% \\
	\bottomrule
	\end{tabular}
	\label{tab:main}
\end{table}

\begin{figure}[htbp]
\centering
\subfigure[]{
\begin{minipage}{0.48\textwidth}
\centering
\includegraphics[width=6cm]{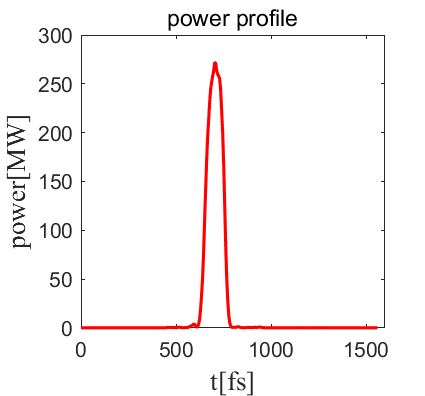}
\end{minipage}
}
\subfigure[]{
\begin{minipage}{0.48\textwidth}
\centering
\includegraphics[width=6cm]{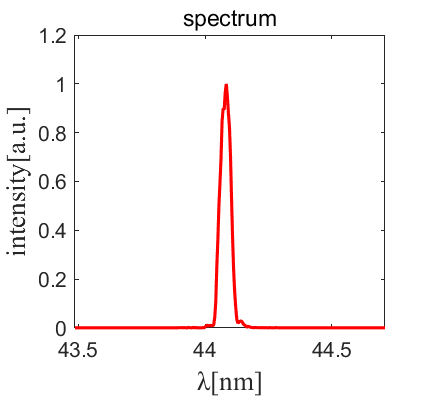}
\end{minipage}
}
\subfigure[]{
\begin{minipage}{0.48\textwidth}
\centering
\includegraphics[width=6cm]{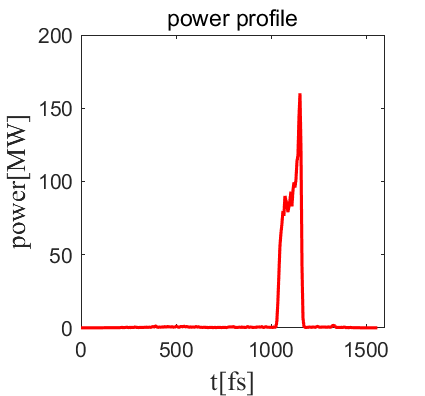}
\end{minipage}
}
\subfigure[]{
\begin{minipage}{0.48\textwidth}
\centering
\includegraphics[width=6cm]{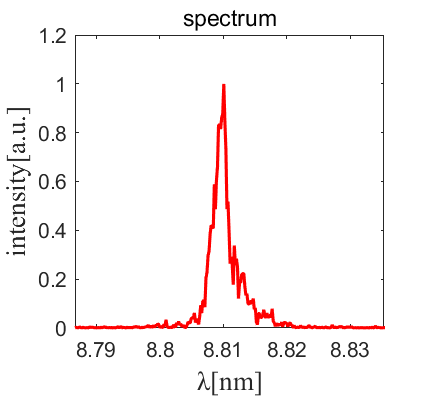}
\end{minipage}
}
\subfigure[]{
\begin{minipage}{0.48\textwidth}
\centering
\includegraphics[width=6cm]{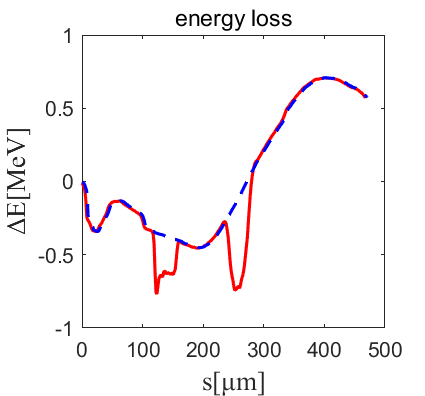}
\end{minipage}
}
\caption{\label{fig}Simulation results given by GENESIS. (a) and (b) are the results of stage-1 while (c) and(d) is of stage-2. (e) shows the the energy loss with(red) and without(blue) lasing.}
\end{figure}

Fig.6(e) clearly shows the energy deviation caused by the wakefield and the wakefield impact on FEL lasing could be estimated from the image. The red line shows the energy loss with two hollows cause by lasing. The right one is caused by lasing in stage-1 and the left one is of stage-2. When the electron beam enters in stage-2 the energy loss at lasing position caused by the wakefield is much more than that in stage-1. Further more, the lasing is more sensitive to energy deviation when it is close to FEL saturation in stage-2. In the meanwhile, the quadratic curvature of the energy loss also brings about an obvious bandwidth broadening to the lasing\cite{shaftan2005high}.

\section{Conclusion}

Wakefields are ineluctable when an electron beam is moving in a pipe with resistive wall or discontinuous structure. The impact on the electron beam due to wakefields in the LINAC section of FEL facility is always ignored because the wakefields are small enough and the impact could be neutralized by tuning the accelerator. However the impact becomes critical in the undulator section especially for cascaded schemes. On the one hand the diagrams of pipes in the undulator section are smaller which means a stronger wakefield, on the other hand it is extremely sensitive in the radiator when it is getting close to FEL saturation. We compared the resistive wall wakefields given by theoretical formulas and ECHO2D, which turned out to be similar, to demonstrate the feasibility of ECHO2D simulation results. On this basis, we measured the wakefield of the overall undulator section of SXFEL test facility. The experimental result after post process agreed well with the simulation result, which made the experiment reliable. In order to illustrate the wakefield impact on the FEL lasing, we used GENESIS to simulate the lasing process at cascaded EEHG-HGHG mode. The experimental and simulation wakefields are adjusted to accommodate the GENESIS simulation. Comparing the lasing simulation results, we saw a minor affect on stage-1 but a critical damage to stage-2 lasing performance. The FEL gain increases along the undulator line while the wakefield also accumulates to degrade the FEL performance. According to the results given in this paper, the pulse energy and peak power of 8.8 nm FEL could decrease by 42\% and 27\% due to the wakefield and could also result in bandwidth broadening. This suggests that the undulator should be well tuned, like taper scheme, according to the wakefield to compensate the loss of FEL performance.

\section{Acknowledgement}

The authors sincerely thank Jiawei Yan for helpful discussions and useful comments. This work was supported by the National Key Research and Development Program of China (2018YFE0103100), the National Natural Science Foundation of China (12125508, 11935020) and Program of Shanghai Academic/Technology Research Leader (21XD1404100).

\bibliography{wake}

\end{document}